\documentclass[%
reprint,
superscriptaddress,
amsmath,amssymb,
aps,
floatfix,
]{revtex4-1}

\usepackage{tabularx}
\usepackage{graphicx}
\usepackage{dcolumn}
\usepackage{bm}
\usepackage{braket}
\usepackage{amsmath}
\usepackage{mathtools}
\usepackage{subfig}
\usepackage{hyperref}

\begin{document}
	
	\raggedbottom
	
	\preprint{APS/123-QED}
	
	\title{High-fidelity State Transfer Between Leaky Quantum Memories}
	
	\author{Daniel Soh}
	\thanks{These authors contributed equally to this work.}
	\affiliation{Sandia National Laboratories, Livermore, California 94550, USA}
	\author{Eric Chatterjee}
	\thanks{These authors contributed equally to this work.}
	\affiliation{Sandia National Laboratories, Livermore, California 94550, USA}
	\author{Matt Eichenfield}
	\affiliation{Sandia National Laboratories, Albuquerque, New Mexico 87123, USA}
	
	\date{\today}
	
	\begin{abstract}
		We derive the optimal analytical quantum-state-transfer control solutions for two disparate quantum memory blocks. Employing the SLH formalism description of quantum network theory, we calculate the full quantum dynamics of system populations, which lead to the optimal solution for the highest quantum fidelity attainable. We show that, for the example where the mechanical modes of two optomechanical oscillators act as the quantum memory blocks, their optical modes and a waveguide channel connecting them can be used to achieve a quantum state transfer fidelity of 96\% with realistic parameters using our derived optimal control solution. The effects of the intrinsic losses and the asymmetries in the physical memory parameters are discussed quantitatively.
	\end{abstract}
	
	\pacs{Valid PACS appear here}
	\maketitle

\section{Introduction}

Many hybrid quantum systems are being explored to enhance the functionality, scalability, and resource-constrained processing power of near-term quantum information processing systems and networks \cite{kurizki2015hybridsystems, wendin2017qip, chu2017quantumacoustics, hong2017optomechanics}. High-fidelity quantum state transfer will enable key network functions such as entanglement distribution \cite{cirac1998quantum} and quantum repeaters \cite{wang2012quantum}. It also allows for various distributed quantum information processing architectures \cite{ritter2012elementary}. High-quality quantum memory serves an important role in these systems by reducing the error correction burden associated with decoherence and loss \cite{lvovsky2009optical, amir2011optomechanical}, as well as coherent quantum storage for processes that require asynchronous qubit operations in algorithms \cite{villalonga2020quantumsupremacyfrontier}. Thus, an ideal quantum memory is a physical system with a long decoherence time and read/write functions enabled via a controllable coupling to intermediary qubits that can propagate between processing blocks and memory blocks or between memory blocks. Regardless of the physical instantiation of quantum memory architecture, the crucial problem to solve is how to configure the time-varying coupling between the storage and intermediary qubits to accomplish optimal quantum state transfer between memory blocks and other elements. 

In recent years, there has been significant progress in understanding how to transfer quantum information to different nodes of a quantum network. For example, Cirac \textit{et al.} considered quantum state transfer between two atomic quantum memories coupled by a photonic channel and optical control fields. They showed that this transfer could be made highly efficient by using a complex but realizable pulse modulation scheme \cite{cirac1998quantum}. Since then, various experiments have demonstrated parts or all of this scheme on optical, microwave, and atomic systems, such as the experimental demonstration over the photon channel \cite{northup2014quantum} and the reversible state transfer between light and an atom \cite{boozer2007reversible}. Some studies have considered the quantum state transfer between flying photons and the cavity storage quanta \cite{palomaki2013coherent}.  More recently, microwave photon quantum state transfer was experimentally demonstrated \cite{kurpiers2018deterministic, axline2018demand}, along with teleportation between optical beams and mechanical modes \cite{hou2016teleportation}. However, the analysis so far does not have the capacity to describe the transfer in the presence of losses and nonidealities, such as leakage into undesired channels. Particularly, an optimal strategy for quantum state transfer between two asymmetric quantum systems with differing energies between the two qubit basis states is completely missing. In a realistic and scalable quantum network, it is highly expected that quantum information will be transferred between nodes that are asymmetric due to the conflicting requirements for performing various quantum operations in each node. 

Here, for the first time, we present the generalized optimal quantum state transfer scheme for two asymmetric quantum systems. Our theory covers quantum memory blocks composed of both identical physical systems and different physical systems with different energy gaps between the two qubit basis states. Our theory is applicable to both the fermionic memory states (such as spin states) and bosonic memory states (such as phonons). We develop a general model of quantum state transfer for open quantum systems with controllable coupling rates, and derive a completely general analytical solution to the optimal control trajectories that accomplishes the highest quantum fidelity obtainable. Knowing the optimal quantum state transfer scheme is crucial to realizing a practical and truly scalable quantum network or hybrid quantum computer. For this, we adopt the emerging SLH formalism description of quantum network theory \cite{combes2017slh, zhang2010direct} to model the entire system and derive the relevant system evolution. Based on the full quantum treatment, we demonstrate that one can determine the optimal quantum state transfer strategy for given system parameters. Our treatment includes realistic nonidealities, i.e., the intrinsic losses, and shows that the optimal solution and the final quantum fidelity are functions of the intrinsic losses. As a concrete example, we will consider two disparate optomechanical quantum memory blocks for optimal quantum state transfer. It will be shown that employing the mechanical mode of an optomechanical oscillator as a quantum memory is tremendously advantageous when implementing such a scheme, since it provides a versatile and controllable coupling mechanism via photon field, along with the desired long decoherence time via the phonon field \cite{stannigel2012optomechanical,meenehan2015phonondecoherence}.

The final goal of this paper is to obtain the optimal time-varying control trajectories that accomplish the highest quantum transfer fidelity. For this, we start with a quantum state transfer model in section II, where the SLH formalism will derive the time evolution of states. Based on this full quantum model, section III solves for the optimal control trajectories for given system parameters, and section IV provides useful numerical examples with realistic optomechanical cavity parameters.

\begin{figure*}[!tb]
	\centering
	\includegraphics[width=0.6\textwidth]{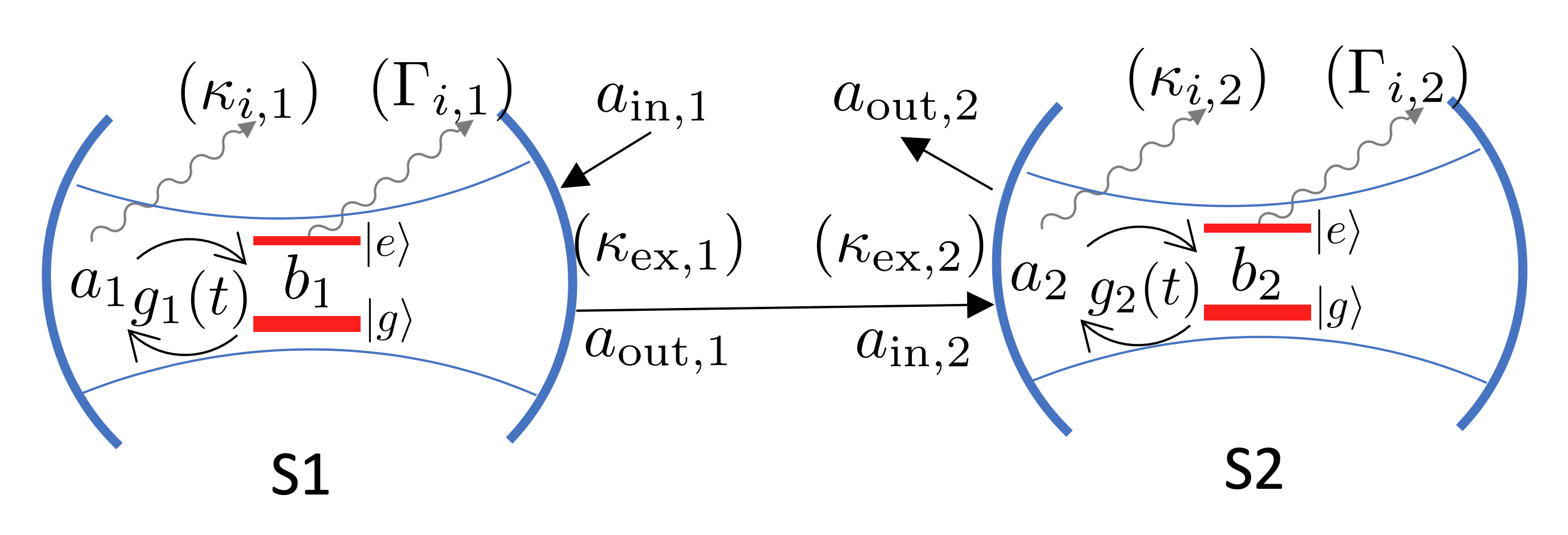}
	\caption{A schematic for general quantum state transfer between two disparate generic memory blocks. ($b_1, b_2$: annihilation operators for the long-decoherence memory states; $a_1, a_2$: annihilation operators for the intermediate states that couple with the input and the output of the memory blocks; $g_1(t), g_2(t)$, the adjustable coupling between the memory states and the intermediate states; $\Gamma_{i,1}, \Gamma_{i,2}$: the intrinsic loss rates of the memory states; $\kappa_{i,1}, \kappa_{i,2}$: the intrinsic loss rates of the intermediate states.)} \label{fig:scheme}
\end{figure*}

\section{Model}

In general, it is useful to build quantum memory from long-decoherence quanta (e.g. phonons), while employing quanta with high propagation velocities (e.g. photons) to transfer information between memory blocks. As such, our goal is to determine how the coupling coefficient between each memory state and the traveling intermediate quantum states connecting the two blocks can be modulated over time such that the transduction fidelity from the first block to the second block via the intermediate states is maximized. Figure~\ref{fig:scheme} depicts the quantum state transfer between two generic quantum memories in block 1 (S1) and block 2 (S2). The aim is to transfer the quantum information initially stored in the long-decoherence memory state of block 1 (S1) to the long-decoherence memory state of block 2 (S2). As for the stored quantum information, we will assume a qubit comprised of the two basis states $\ket{g}$ and $\ket{e}$, representing vacuum and singly-excited states, respectively. Intermediate quantum states in memory blocks,  which couple with the input and the output of the memory blocks, mediate the quantum information to and from the memory states when the coupling mechanism ($g_1 (t), g_2(t)$) is turned on. 

To start the analysis, we establish a Hilbert space composed of a tensor product of the memory state for S1, the memory state for S2, the intermediate quantum state for S1, and the intermediate quantum state for S2, represented respectively by the annihilation (creation) operators $b_1^{(\dag)}$, $b_2^{(\dag)}$, $a_1^{(\dag)}$, and $a_2^{(\dag)}$. The S1 memory is initially in an arbitrary superposition of ground and excited states, corresponding to the coefficients $c_g$ and $c_e$. Therefore, the initial state of the overall system can be expressed as follows:
\begin{align} \label{eq: initial wavefunction}
    \ket{\Psi(t_i)} &=(c_g \ket{g} + c_e \ket{e})\ket{g} \ket{00} \nonumber \\
    &= c_g\ket{gg}\ket{00} + c_e\ket{eg}\ket{00}.
\end{align}
The two slots of the first ket are the memory states of S1 and S2, respectively, and those of the second ket are the intermediate states of S1 and S2. Here, we assume that the memory states are qubits. Our goal is to faithfully transmit the excitation in $b_1$ to $b_2$, such that the final state becomes
\begin{equation} 
    \ket{\Psi(t_f)} = c_g\ket{gg}\ket{00} + c_e\ket{ge}\ket{00}.
\end{equation}
Note that as long as the system starts with 0 or 1 excitation in the S1 memory mode and vacuum in all other modes, energy conservation ensures that the total number of excitations in the system at any given time can never exceed 1, thus restricting the Hilbert space to 5 dimensions, spanned by $\ket{gg}\ket{00}$, $\ket{eg}\ket{00}$, $\ket{gg}\ket{10}$, $\ket{gg}\ket{01}$, and $\ket{ge}\ket{00}$. This ensures that the probability that the $b_2$ mode will be multiply-occupied at the end of the transfer process is always zero.
\par
Next, we determine the Hamiltonian governing this system. Given a time-varying interaction $g(t)$ between the memory state and the intermediate quantum state, we consider a class of memory blocks that couples the mediating cavity field and the long-term memory field via the following interaction Hamiltonian:
\begin{equation}
    H_\mathrm{int} = -\hbar g(t) (a^{\dag}b + ab^{\dag}).
\end{equation}
In fact, this model covers a vast majority of quantum memory configurations since in most quantum memories the read and write function is accomplished via such a beam-splitter coupling Hamiltonian between the mediating cavity field (photons or phonons) and the long-term memory state. We note that this model covers both the direct coupling such as dipole coupling between a cavity photon field and the spin memory state, and the parametric coupling between the photon field and the phonon memory states in an optomechanical system. For the example of optomechanical oscillators as the memory block, $g(t) = g_0 \sqrt{\bar{n} (t)}$ with the single-photon coupling coefficient $g_0 = G x_\mathrm{zpf}$. $G = - \partial \omega_\mathrm{cav} / \partial x$ is the slope of photonic resonance frequency over the displacement $x$ of the mechanical system, and $x_\mathrm{zpf}$ is the zero-point-fluctuation parameter of the mechanical system. Varying the control $g(t)$ is accomplished through varying the photon population $\bar{n}(t)$ at a red-detuned driving frequency, while varying $g_0$ is also possible as we will explain later. For another example of the neutral atoms as the memory block, $g(t)$ is the Rabi oscillation frequency, a function of dipole moment and an adjustable driving electromagnetic field amplitude. 

Our system consists of two blocks $i = 1,2$, with the following Hamiltonian for each:
\begin{equation}
    H^{(i)}_\mathrm{int} = -\hbar g_i(t) (a_i^{\dag}b_i + a_ib_i^{\dag}).
\end{equation}
It is essential to consider the various loss channels from these blocks, which we will model using Lindblad operators. In particular, the state transfer process \textit{requires} loss of the intermediate state from S1, so that the resulting propagating wavepackets can serve as the input to S2. We will label the rates for this coherent extrinsic loss process as $\kappa_\mathrm{ex}$ and $\kappa'_\mathrm{ex}$ for S1 and S2, respectively. In addition, we will need to account for the incoherent intrinsic losses from the blocks. For the respective blocks, we label the loss rates for the intermediate quanta as $\kappa_{i,1}$ and $\kappa_{i,2}$, and the decay rates for the memory storage quanta as $\Gamma_{i,1}$ and $\Gamma_{i,2}$. The Lindblad operators for the two blocks are thus expressed in the following manner:
\begin{equation}
    L_1 =
    \begin{pmatrix}
    \sqrt{\kappa_{\mathrm{ex},1}} a_1 \\
    \sqrt{\kappa_{i,1}} a_1 \\
    \sqrt{\Gamma_{i,1}} b_1
    \end{pmatrix}, \quad
    L_2 =
    \begin{pmatrix}
    \sqrt{\kappa_{\mathrm{ex},2}} a_2 \\
    \sqrt{\kappa_{i,2}} a_2 \\
    \sqrt{\Gamma_{i,2}} b_2
    \end{pmatrix}.
\end{equation}
We now consider the overall system consisting of the two blocks. Based on the SLH formalism \cite{combes2017slh}, we establish the following input-output relationships for the blocks:
\begin{equation}
    a_\mathrm{out,n} = a_\mathrm{in,n} + \sqrt{\kappa_{\mathrm{ex},n}} a_n,
\end{equation}
where $n = 1,2$. The input of the second block is identically the output of the first block, except for an appropriate time delay. One can eliminate the time delay effectively from all formula by introducing the \textit{time delayed} operators for the operators in S2 \cite{cirac1998quantum}. Applying this time adjustment, one can set $a_\mathrm{in,2} (t) = a_\mathrm{out,1} (t)$ with the implicit understanding of $a_\mathrm{in,2} (t) \rightarrow \tilde{a}_\mathrm{in,2} (t) (= a_\mathrm{in,2} (t - \tau))$ where $\tau$ is the time delay. The output operator for S2, representing a loss channel for the system, then takes the form of a superposition of the two intermediate state annihilation operators and the bath input operator for S1:
\begin{equation}
    a_\mathrm{out,2} = \sqrt{\kappa_{\mathrm{ex},1}} a_1 + \sqrt{\kappa_{\mathrm{ex},2}} a_2 + a_\mathrm{in,1}.
\end{equation}
We assume that the bath temperature is sufficiently low and thus, the input fields are in vacuum states. Then, $a_\mathrm{in,1}$ returns a null value when applied on the system state. According to the series connection rule of the SLH formalism, the connected system's Hamiltonian $H_T$ and Lindbladian $L_T$ take the following form:
\begin{align} \label{eq: H_T}
    H_T &= -\hbar g_1(t) (a_1^{\dag}b_1 + a_1b_1^{\dag}) - \hbar g_2(t) (a_2^{\dag}b_2 + a_2b_2^{\dag}) \nonumber \\
    &\quad + \frac{i\hbar}{2} \sqrt{\kappa_{\mathrm{ex},1}\kappa_{\mathrm{ex},2}} (a_1^{\dag}a_2 - a_1a_2^{\dag}), \\
    L_T &=
    \begin{pmatrix}
    \sqrt{\kappa_{i,1}} a_1 \\
    \sqrt{\Gamma_{i,1}} b_1 \\
    \sqrt{\kappa_{\mathrm{ex},1}} a_1 + \sqrt{\kappa_{\mathrm{ex},2}} a_2 \\
    \sqrt{\kappa_{i,2}} a_2 \\
    \sqrt{\Gamma_{i,2}} b_2
    \end{pmatrix}.
\end{align}
One way to physically explain the above result is as follows: the sub-processes corresponding to extrinsic loss from the individual blocks are split into a part representing coherent coupling between blocks, which we incorporate in the composite Hamiltonian $H_T$, and a part represented by $a_\mathrm{out,2}$ corresponding to system loss due to reflection of the propagating wavepacket from the second block, which enters into the Lindbladian $L_T$. 

\section{Analytical solution}
It is useful to incorporate the Lindblad operators into the Hamiltonian in order to model the non-unitary time evolution of the wavefunction in the basis of the excited states. In general, for a set of loss channels labeled $n$, the effective Hamiltonian takes the following form \cite{cirac1998quantum}:
\begin{equation}
    H^\mathrm{eff} = H - \sum_{n} \frac{i}{2} L_n^{\dag} L_n.
\end{equation}
Since coherent reflection $a_\mathrm{out,2}$ serves as the dominant loss channel, we start by expanding the Hamiltonian to incorporate the corresponding Lindblad operator:
\begin{align}
\begin{split}
    H' &= H_T - \frac{i\hbar}{2} \Big(\sqrt{\kappa_{\mathrm{ex},1}} a_1^{\dag} + \sqrt{\kappa_{\mathrm{ex},2}} a_2^{\dag}\Big) \Big(\sqrt{\kappa_{\mathrm{ex},1}} a_1 \\
    &\quad + \sqrt{\kappa_{\mathrm{ex},2}} a_2\Big).
\end{split}
\end{align}
By default, this effective Hamiltonian is non-Hermitian for the overall Hilbert space. However, we can design the couplings $g_1(t)$ and $g_2(t)$ such that the system remains in a dark state throughout the transfer process so that the system states evolve coherently \cite{cirac1998quantum}. 

We first ignore the intrinsic losses. Defining the dimensionless constant $\epsilon$ such that $\kappa_{\mathrm{ex},2} = \epsilon \kappa_{\mathrm{ex},1}$, we derive the dark mode (whose annihilation operator is represented by $a_d$) by superposing $a_1$ and $a_2$ for the intermediate modes such that applying the annihilation operator $a_d = a_1 + \sqrt{\epsilon} a_2$ onto the dark state returns a null value:
\begin{equation}
    \ket{\mathrm{dark}} = \frac{1}{\sqrt{1+\epsilon}} \Big(\ket{01} - \sqrt{\epsilon}\ket{10}\Big).
\end{equation}
We note that this dark state produces zero quanta in the S2 output channel (no information leakage) as the connected system's Lindblad operator produces a null result on this state. 

It is also necessary to define the bright mode (whose annihilation operator is represented by $a_b$) as orthogonal to the dark mode:
\begin{equation}
    \ket{\mathrm{bright}} = \frac{1}{\sqrt{1+\epsilon}} \Big(\sqrt{\epsilon}\ket{01} + \ket{10}\Big).
\end{equation}
The bright state produces nonzero S2 output quanta, losing quantum information from the connected system, and therefore, we want to avoid such a bright state in order to maximize the quantum state transfer fidelity.

For the kernel of the $a_b^{\dag}a_b$ operator (i.e., the reduced Hilbert space where applying $a_b^\dagger a_b$ produces null for all elements), $H'$ becomes Hermitian because the system does not decohere:
\begin{widetext}
\begin{align}
    \begin{split}
    H' &= \hbar g_1(t) \frac{\sqrt{\epsilon}}{\sqrt{1+\epsilon}} (a_d^{\dag}b_1 + a_db_1^{\dag}) - \hbar g_1(t) \frac{1}{\sqrt{1+\epsilon}} (a_b^{\dag}b_1 + a_bb_1^{\dag}) - \hbar g_2(t) \frac{1}{\sqrt{1+\epsilon}} (a_d^{\dag}b_2 + a_db_2^{\dag}) \\
    &\quad - \hbar g_2(t) \frac{\sqrt{\epsilon}}{\sqrt{1+\epsilon}} (a_b^{\dag}b_2 + a_bb_2^{\dag}) + i\hbar \frac{\kappa_{\mathrm{ex},1}}{2} \sqrt{\epsilon} (a_b^{\dag}a_d - a_ba_d^{\dag}) - i\hbar \frac{\kappa_{\mathrm{ex},1}}{2} (1+\epsilon) a_b^{\dag} a_b.
    \end{split}
\end{align}
\end{widetext}
Given the initial composite state described in Eq.~\eqref{eq: initial wavefunction}, the time-dependent composite state can be expressed in terms of generic coefficients $\alpha_1(t)$, $\alpha_2(t)$, and $\beta_{\alpha}(t)$:
\begin{align} \label{eq: time-dependent wavefunction}
\begin{split}
\ket{\Psi(t)} &= c_g\ket{gg}\ket{00} + c_e\Big[\alpha_1(t)\ket{eg}\ket{00} + \alpha_2(t)\ket{ge}\ket{00} \\
&\quad + i\beta_{\alpha}(t)\ket{gg}\ket{\mathrm{dark}}\Big].
\end{split}
\end{align}

Next, we introduce the intrinsic, decohering loss processes to the effective Hamiltonian. Since we design both blocks such that these intrinsic loss rates are equally minimal, we assume that $\kappa_i = \kappa_{i,1} = \kappa_{i,2}$ and $\Gamma_i = \Gamma_{i,1} = \Gamma_{i,2}$ for simplicity. However, we note that this simplifying assumption can be removed, and it is straightforward to derive the general result in the case of asymmetric intrinsic losses. This results in the effective Hamiltonian:
\begin{align}
    H^\mathrm{eff}_T = H' - i\hbar \frac{\kappa_i}{2} (a_b^{\dag}a_b + a_d^{\dag}a_d) - i\hbar \frac{\Gamma_i}{2} (b_1^{\dag}b_1 + b_2^{\dag}b_2).
\end{align}
Now that we have decoherence in the system, by breaking the Hermiticity of the Hamiltonian, the intrinsic losses introduce non-unitarity to the time evolution in the Hilbert space of the excited states, such that $|\alpha_1(t)|^2 + |\alpha_2(t)|^2 + |\beta_{\alpha}(t)|^2 < 1$ for any time $t > t_i$. Conceptually, the decline of the normalization value over time corresponds to the increasing probability that the system has collapsed to the vacuum state $\ket{gg}\ket{00}$, which effectively increases the value of $|c_g|^2$. 
\par
Applying the Schr\"{o}dinger equation, we obtain the dynamically coupled differential equations:
\begin{align}
    \label{eq: dot alpha_1 asym}
    \dot{\alpha_1}(t) &= \frac{\sqrt{\epsilon}}{\sqrt{1+\epsilon}} g_1(t) \beta_{\alpha}(t) - \frac{\Gamma_i}{2} \alpha_1(t), \\
    \label{eq: dot alpha_2 asym}
    \dot{\alpha_2}(t) &= -\frac{1}{\sqrt{1+\epsilon}} g_2(t) \beta_{\alpha}(t) - \frac{\Gamma_i}{2} \alpha_2(t), \\
    \label{eq: dot beta_alpha asym}
    \dot{\beta_{\alpha}}(t) &= \frac{1}{\sqrt{1+\epsilon}} g_2(t) \alpha_2(t) - \frac{\sqrt{\epsilon}}{\sqrt{1+\epsilon}} g_1(t) \alpha_1(t) - \frac{\kappa_i}{2} \beta_{\alpha}(t).
\end{align}
To prevent the system from entering the bright mode, we constrict the range of coupling profiles $g_1 (t)$ and $g_2(t)$ such that:
\begin{equation}
    \bra{gg,\mathrm{bright}} H^\mathrm{eff}_T \ket{\Psi(t)} = 0.
\end{equation}
This yields a fourth equation for the coefficients:
\begin{equation} \label{eq: dot beta_s asym}
    0 = \frac{\sqrt{\epsilon}}{\sqrt{1+\epsilon}} g_2(t) \alpha_2(t) + \frac{1}{\sqrt{1+\epsilon}} g_1(t) \alpha_1(t) + \frac{\kappa_{\mathrm{ex},1}\sqrt{\epsilon}}{2} \beta_{\alpha}(t).
\end{equation}
The supplemental material demonstrates the method for solving the time-evolution of the coefficients, defining $t = 0$ as the halfway-point of the process. To reduce the range of possible coupling profiles, we set $g_1(t)$ and $g_2(t)$ as constant for $t \ge 0$ and $t \le 0$, respectively, following Cirac \textit{et al.} \cite{cirac1998quantum}. The reasoning is that in the second (first) half of the process, the excited population will largely be in S2 (S1), and therefore changes in the coupling for S1 (S2) will not substantially alter the state populations. It is worth noting that Xu \textit{et al.} \cite{xu2016controllable} used a linear combination of Gaussian functions to construct the coupling profiles but also obtained an approximately flat shape for the driving field for Segment 1 (2) when the excited population was largely in Segment 2 (1). For $t \ge 0$, we derive the following expressions (see the supplemental material):
\begin{widetext}
\begin{align} \label{eq: alpha_1(t) t>0 asym}
    \begin{split}
    \alpha_1(t) &= \frac{e^{-\frac{1}{4}(\kappa_{\mathrm{ex},1} + \kappa_i + \Gamma_i)t}}{C} \bigg(\alpha_1(0)C\cosh{\bigg(\frac{C}{4}t\bigg)} + B'_2\sinh{\bigg(\frac{C}{4}t\bigg)} \bigg).
    \end{split} \\
    \label{eq: beta_alpha(t) t>0 asym}
    \begin{split}
    \beta_{\alpha}(t) &= \frac{e^{-\frac{1}{4}(\kappa_{\mathrm{ex},1} + \kappa_i + \Gamma_i)t}}{C} \bigg(\beta_{\alpha}(0)C\cosh{\bigg(\frac{C}{4}t\bigg)} - B'_1\sinh{\bigg(\frac{C}{4}t\bigg)} \bigg).
    \end{split}
\end{align}
\begin{align} \label{eq: alpha_2(t) t>0 asym}
    \begin{split}
    \alpha_2(t)^2 &= \frac{e^{-\frac{1}{2}(\kappa_{\mathrm{ex},1} + \kappa_i + \Gamma_i)t}}{16g_1(0)^2} \Bigg(\bigg(\kappa_{\mathrm{ex},1} \beta_{\alpha}(0)^2 + \frac{2g_1(0)}{\sqrt{\epsilon(1+\epsilon)}} \alpha_1(0) \beta_{\alpha}(0)\bigg) A_1(t) + \bigg(2\kappa_{\mathrm{ex},1} B'_1 \beta_{\alpha}(0) \\
    &\quad + \frac{2g_1(0)}{\sqrt{\epsilon(1+\epsilon)}} \Big(\alpha_1(0)B'_1 - \beta_{\alpha}(0)B'_2\Big)\bigg) A_2(t) + \bigg(\kappa_{\mathrm{ex},1}B_1^{'2} - \frac{2g_1(0)}{\sqrt{\epsilon(1+\epsilon)}} B'_1 B'_2\bigg) A_3(t) \Bigg) + Ge^{-\Gamma_i t}.
    \end{split}
\end{align}
\end{widetext}
Here, $C = \sqrt{(\Gamma_i - \kappa_{\mathrm{ex},1} - \kappa_i)^2 - 16g_1(0)^2}$. Note that as $t$ increases, $\alpha_1(t)$ and $\beta_{\alpha}(t)$ converge to zero as expected, since the system should eventually either transition fully to the $b_2$ memory mode or decay to the vacuum state. The functions $A_1(t)$, $A_2(t)$, and $A_3(t)$ are defined as follows:
\begin{widetext}
\begin{equation}
A_{1,3}(t) = \frac{\mp C^2 - C(\Gamma_i - \kappa_i - \kappa_{\mathrm{ex},1})\sinh{\Big(\frac{C}{2}t\Big)} + (\Gamma_i - \kappa_i - \kappa_{\mathrm{ex},1})^2\Big(\cosh{\Big(\frac{C}{2}t\Big)} \pm 1\Big)}{C^{1 \mp 1}(\Gamma_i - \kappa_i - \kappa_{\mathrm{ex},1})}.
\end{equation}
\begin{equation}
A_2(t) = \frac{-(\Gamma_i - \kappa_i - \kappa_{\mathrm{ex},1})\sinh{\Big(\frac{C}{2}t\Big)} + C\cosh{\Big(\frac{C}{2}t\Big)}}{C}.
\end{equation}
\end{widetext}
and the constants $B'_1$ and $B'_2$ are the following:
\begin{align} \label{eq: B'_1}
    B'_1 &= 4\sqrt{\frac{1+\epsilon}{\epsilon}}g_1(0)\alpha_1(0) + \beta_{\alpha}(0)(\kappa_{\mathrm{ex},1} + \kappa_i - \Gamma_i), \\
    \label{eq: B'_2}
    B'_2 &= 4 \sqrt{\frac{\epsilon}{1+\epsilon}} g_1(0)\beta_{\alpha}(0) + \alpha_1(0)(\kappa_{\mathrm{ex},1} + \kappa_i - \Gamma_i).
\end{align}
Regarding the term $Ge^{-\Gamma_i t}$, it is useful to note that $\Gamma_i \ll \kappa_{\mathrm{ex},1},\kappa_i$ due to the extremely long memory decoherence time. Therefore, over the timescale of the transfer process, $e^{-\Gamma_i t}$ will approximately equal 1, resulting in $G \approx \alpha_2^2(t_f)$, which represents the fidelity of quantum state transfer.
\par
For $t \le 0$, the coefficient solutions take a similar form (see the supplemental section), with the replacements $\alpha_1 \leftrightarrow \alpha_2$, $\kappa_{\mathrm{ex},1} \rightarrow -\epsilon\kappa_{\mathrm{ex},1}$, $C \rightarrow D'$ (where $D' = \sqrt{(\Gamma_i + \epsilon\kappa_{\mathrm{ex},1} - \kappa_i)^2 - 16g_2(0)^2}$), $2g_1(0)/\sqrt{\epsilon(1+\epsilon)} \rightarrow -2\epsilon g_2(0)/\sqrt{1+\epsilon}$ in the coefficient for $\alpha_1(t)^2$, $G \rightarrow G'$, as well as $B'_1 \rightarrow B'_3$ and $B'_2 \rightarrow B'_4$, where:
\begin{align} \label{eq: B'_3}
    B'_3 &= -4\Big(\sqrt{1+\epsilon}\Big)g_2(0)\alpha_2(0) + \beta_{\alpha}(0)(-\epsilon\kappa_{\mathrm{ex},1} + \kappa_i - \Gamma_i), \\
    \label{eq: B'_4}
    B'_4 &= -\frac{4g_2(0)}{\sqrt{1+\epsilon}}\beta_{\alpha}(0) + \alpha_2(0)(-\epsilon\kappa_{\mathrm{ex},1} + \kappa_i - \Gamma_i).
\end{align}
As previously mentioned, $e^{-\Gamma_i t} \approx 1$ for all times $t_i \le t \le t_f$. Therefore, $G' \approx \alpha_1(t_i)^2$, and since the system starts out in the $b_1$ memory mode, this implies that $G' \approx 1$.
\par
Having found solutions for the time-evolution of the coefficients $\alpha_1(t)$, $\alpha_2(t)$, and $\beta_{\alpha}(t)$ in terms of their values at $t = 0$, we can now use these functions along with the known constants $g_1(0)$ and $g_2(0)$ to determine the optimal coupling profile functions. 
\begin{table*}[t]
    \centering
    \resizebox{2\columnwidth}{!}{%
    \begin{tabular}{c|c|c}
         & $t \le 0$ & $t \ge 0$\\ &&\\[-2mm] \hline
    &&\\[-2mm]
    $g_1(t)$ & $-\dfrac{1}{\alpha_1 (t)}\left(\dfrac{\kappa_{\mathrm{ex},1}\sqrt{\epsilon(1+\epsilon)}}{2} \beta_{\alpha}(t) + \sqrt{\epsilon} g_2(0) \alpha_2(t)\right)$ & $g_1(0)$\\ 
    &&\\[-2mm]\hline
    &&\\[-2mm]
    $g_2(t)$ & $g_2(0)$ & $-\dfrac{1}{\alpha_2 (t)}\left(\dfrac{\kappa_{\mathrm{ex},1}\sqrt{1+\epsilon}}{2} \beta_{\alpha}(t) + \dfrac{1}{\sqrt{\epsilon}} g_1(0) \alpha_1(t)\right)$ \\
    \end{tabular} %
    }
    \caption{Optimal coupling $g_1(t)$ and $g_2(t)$ for $t \le 0$ and $t \ge 0$ producing the highest quantum state transfer fidelity attainable.}
    \label{tab:coupling coefficients}
\end{table*}
The expressions for $g_1(t)$ and $g_2(t)$ are shown in Table~\ref{tab:coupling coefficients} (for detailed derivation, see the supplemental material).

\section{Numerical methods}

\begin{figure}[!tb]
	\centering
	\includegraphics[width=\linewidth]{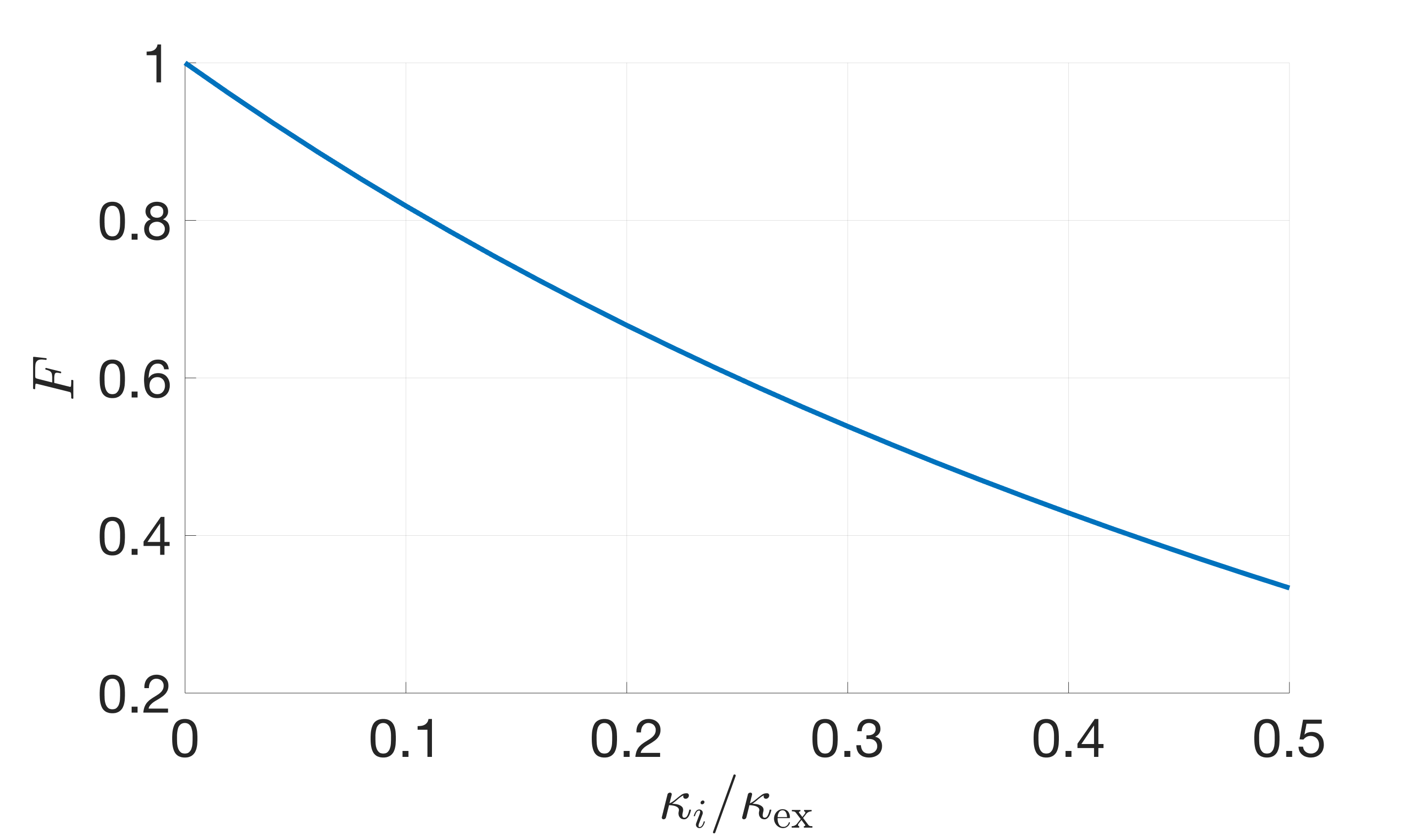}
	\caption{Optimal fidelity as a function of ratio of intrinsic loss rate to output coupling rate, assuming equal output coupling rates for the two blocks, i.e. $\kappa_\mathrm{ex} = \kappa_{\mathrm{ex},1} = \kappa_{\mathrm{ex},2}$.}
	\label{fig:fidelity2d}
\end{figure}

In order to determine numerical solutions for the coupling profiles, we used Matlab-based computation to calculate the valid sets of $(\alpha_1(0),\alpha_2(0),\beta_{\alpha}(0))$. Due to the extremely long memory decoherence time, we can generally set $\Gamma_i \approx 0$. However, the intrinsic intermediate state loss will measurably degrade the fidelity, as depicted in Figure~\ref{fig:fidelity2d}.
The inverse correlation between the fidelity and the ratio of intrinsic loss to output coupling rate can be interpreted as follows: The output coupling corresponds to the rate of transfer from S1 to S2. A higher rate shortens the time over which the excitation is in the intermediate form, thus reducing the probability that the excitation decays due to intrinsic loss.
\par
\begin{figure}[!tb]
	\centering
	\includegraphics[width=\linewidth]{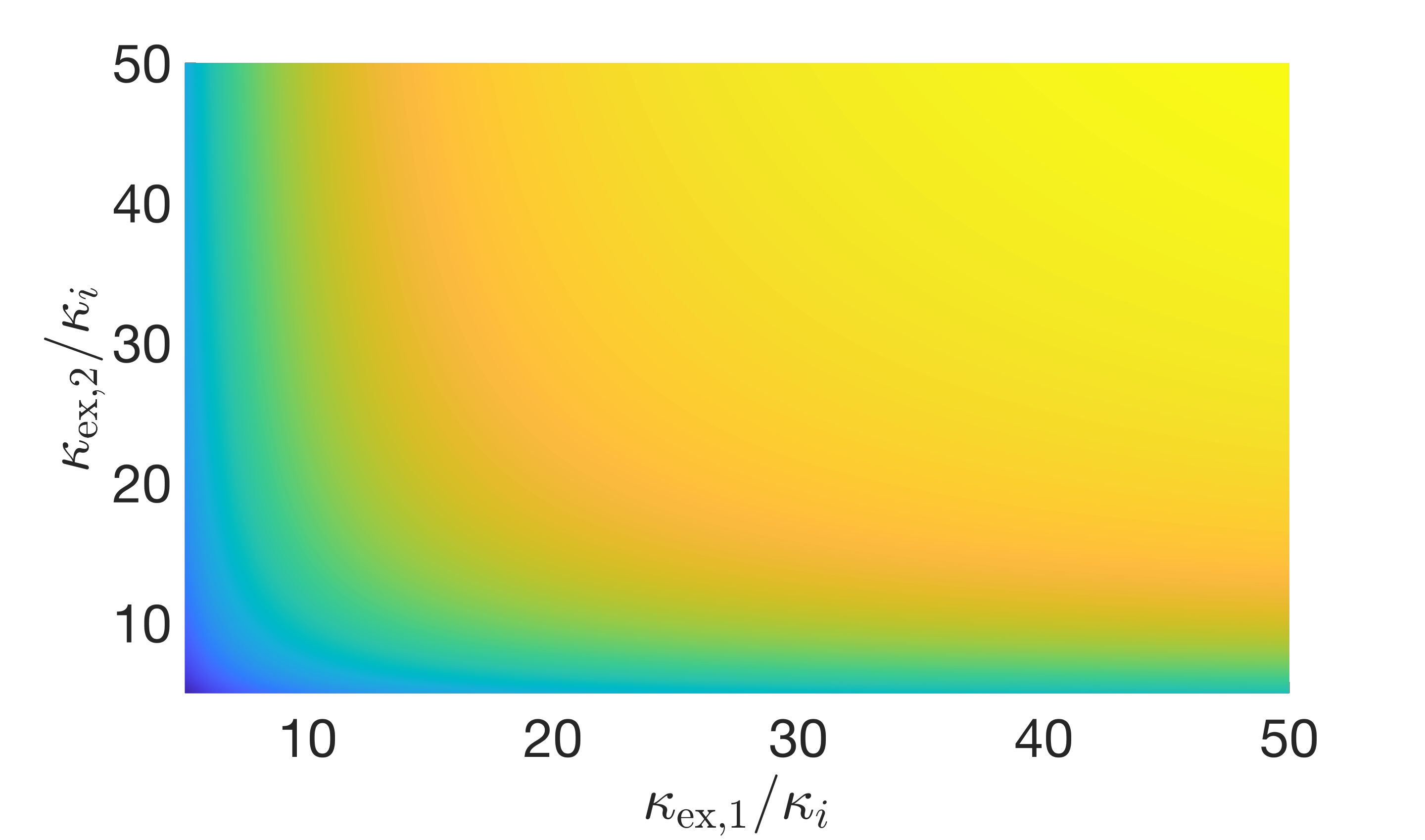}
	\caption{Optimal fidelity as a function of output coupling to intrinsic loss ratio for S1 and S2.}
	\label{fig:fidelity3d}
\end{figure}

 
It is also useful to consider the effect of asymmetry between the blocks. Figure~\ref{fig:fidelity3d} depicts the optimal fidelity as a function of the output coupling rates of the two blocks, with the intrinsic loss rates set at the minimal value, i.e. $\kappa_i = \kappa_{i,1} = \kappa_{i,2}$.
As expected, the fidelity declines if either of the extrinsic loss rates is reduced, approximately fitting an exponential profile, which follows from the nearly exponential variation of the fidelity with the reciprocal of the extrinsic loss as shown in Figure~\ref{fig:fidelity2d}.
\par
\begin{figure*}[!tb]
	\centering
	\subfloat[State population vs. time.]{\includegraphics[width=0.48\textwidth]{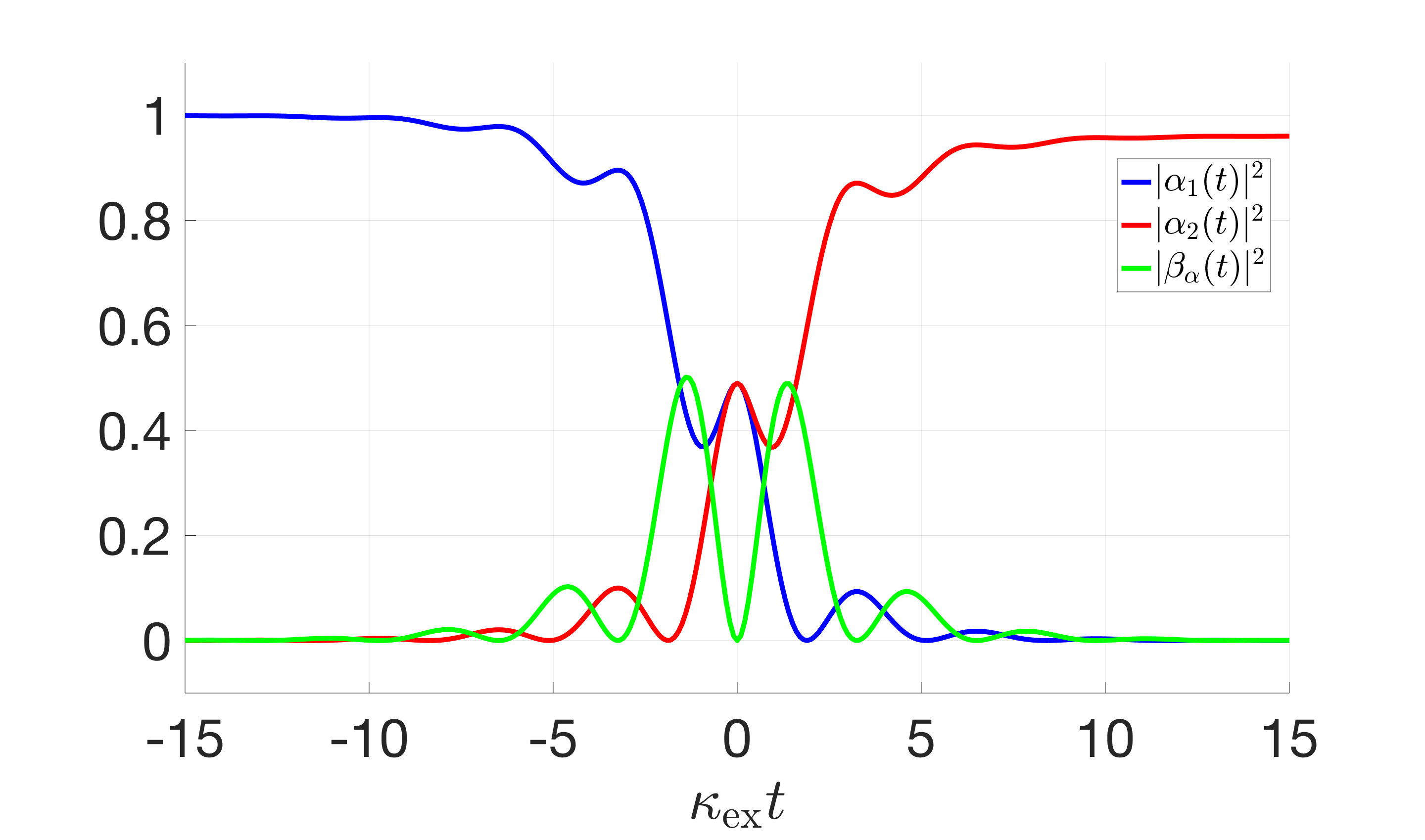}}
	\subfloat[Optimal control profiles vs. time.]{\includegraphics[width=0.48\textwidth]{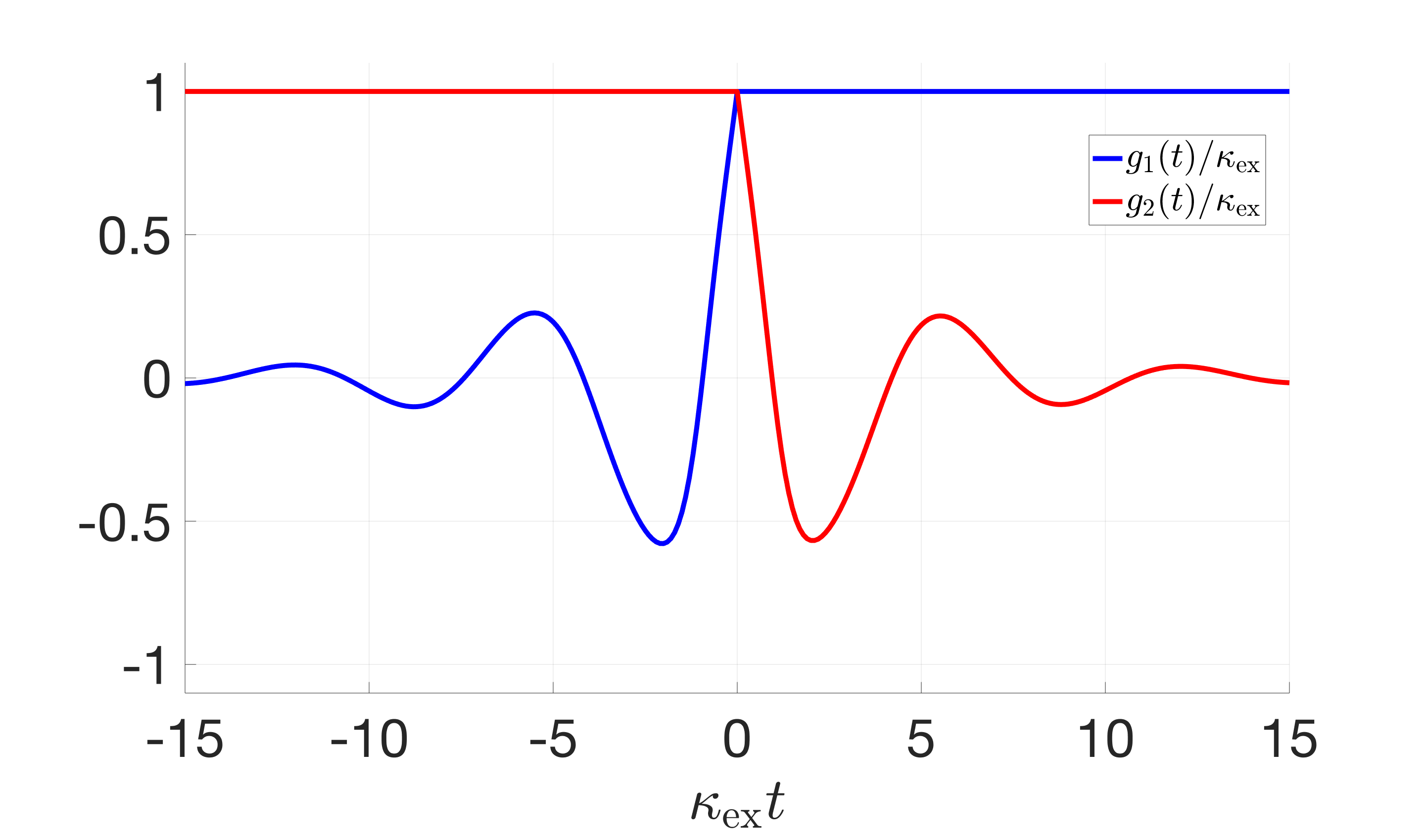}}
	\caption{Evolution of the state populations $|\alpha_1(t)|^2$, $|\alpha_2(t)|^2$, and $|\beta_{\alpha}(t)|^2$ (a) and optimal pulse profiles $g_1(t)$ and $g_2(t)$ (b) over time for $\kappa_\mathrm{ex} = \kappa_{\mathrm{ex},1} = \kappa_{\mathrm{ex},2} = 2\pi \times 5 \textrm{ GHz}$, $\kappa_i = 2\pi \times 100 \textrm{ MHz}$, $g_1(0) = \kappa_\mathrm{ex} = g_2(0)$, $\alpha_1(0) = 0.7$, $\alpha_2(0) = -0.7$, and $\beta_{\alpha}(0) = 0$.}
    \label{fig:376i}
\end{figure*}
As a useful example, we consider empirical values for optomechanical oscillator blocks. A high-quality optomechanical oscillator features a mechanical damping rate of $\Gamma_i \lesssim 2 \pi \times 5 \textrm{ kHz}$ and an intrinsic photon loss rate of $\kappa_i \approx 2 \pi \times 100 \textrm{ MHz}$, along with an upper bound for $2 \pi \times 5 \textrm{ GHz}$ for the output coupling rates $\kappa_{\mathrm{ex},1}$ and $\kappa_{\mathrm{ex},2}$. The very low value for the mechanical damping rate meets our requirement that the memory decoherence be negligible over the timescale of the process, and we therefore set $\Gamma_i \approx 0$. Having shown that the fidelity is highest for maximal output coupling, we also set $\kappa_{\mathrm{ex},1} = 2\pi \times 5 \textrm{ GHz} = \kappa_{\mathrm{ex},2}$. For these inputs, our calculations yield an optimal fidelity of about 96\%, independent of $g_1(0)$ or $g_2(0)$. For the choice of $g_1(0) = \kappa_{\mathrm{ex},1} = g_2(0)$, $\alpha_1(0) = 0.7$, $\alpha_2(0) = -0.7$, and $\beta_{\alpha}(0) = 0$, Figure~\ref{fig:376i} depicts the time-evolution of the state populations $|\alpha_1(t)|^2$, $|\alpha_2(t)|^2$, and $|\beta_{\alpha}(t)|^2$, and the corresponding pulse profiles. Note that the total population degrades due to intrinsic loss, with the fidelity represented by the final population $|\alpha_2(t_f)|^2 \approx 0.96$. For the given $t = 0$ values, it is also evident that in the first half of the process, half of the population is transferred to the final phonon mode, which is then repeated with the other half of the population. This indicates that the process is close to time-reversible (though full time-reversibility is prevented by intrinsic loss), corresponding to the observation that $g_1(t)$ and $g_2(-t)$ closely match each other.
\par 
\begin{figure*}[!tb]
	\centering
	\subfloat[State population vs. time.]{\includegraphics[width=0.48\textwidth]{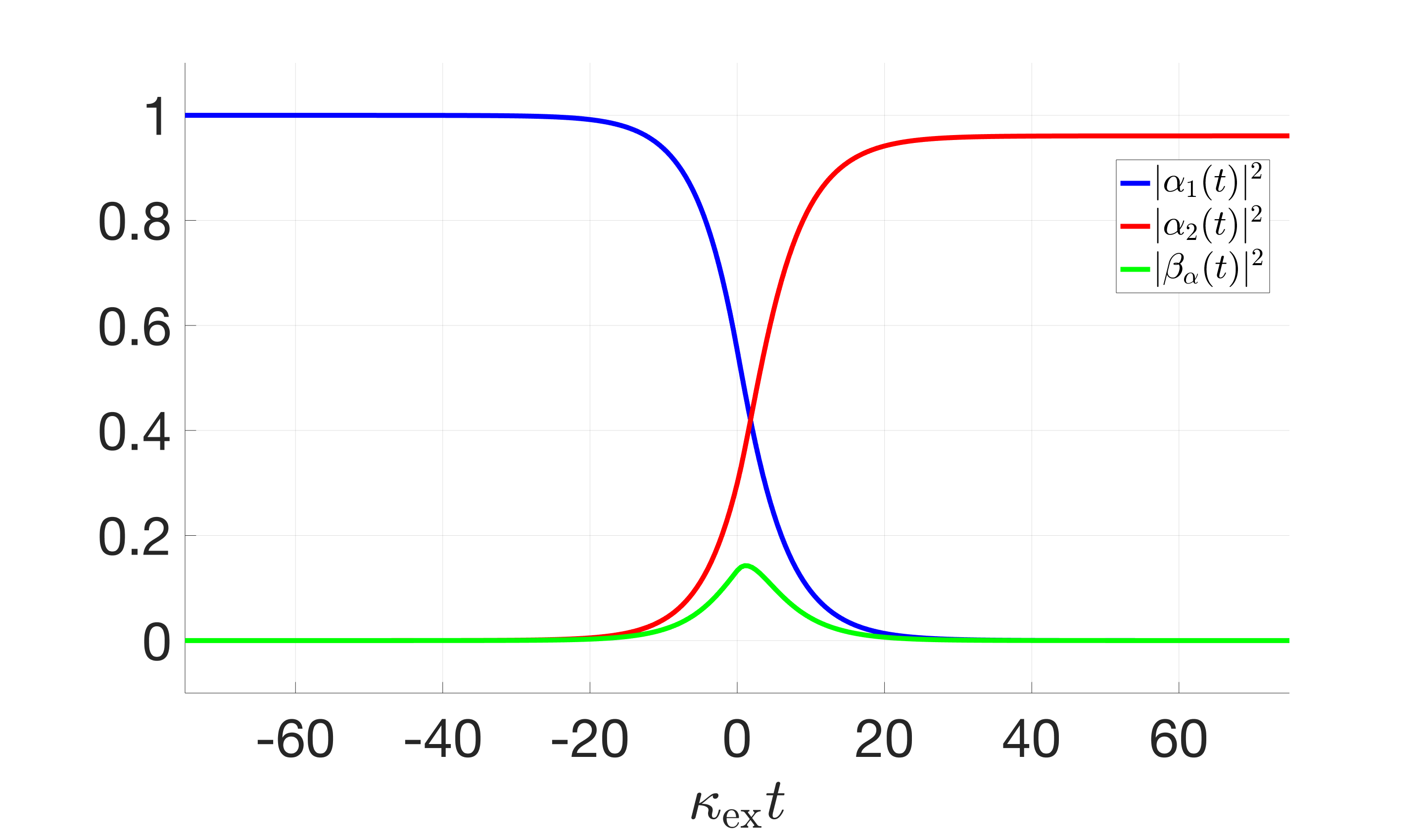}}
	\subfloat[Optimal control profile vs. time.]{\includegraphics[width=0.48\textwidth]{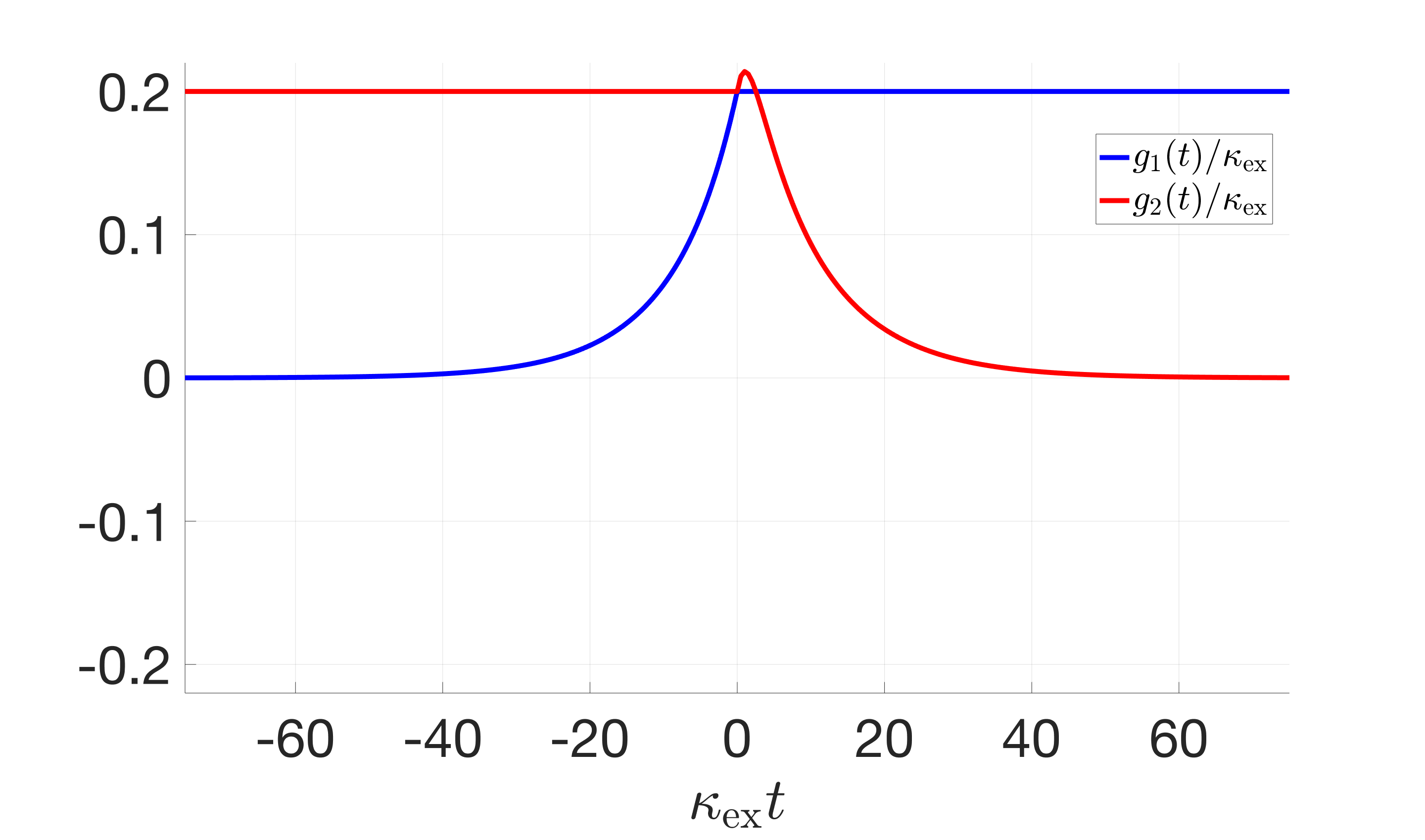}}
	\caption{Evolution of the state populations $|\alpha_1(t)|^2$, $|\alpha_2(t)|^2$, and $|\beta_{\alpha}(t)|^2$ (a) and optimal pulse profiles $g_1(t)$ and $g_2(t)$ (b) over time for $\kappa_\mathrm{ex} = \kappa_{\mathrm{ex},1} = \kappa_{\mathrm{ex},2} = 2\pi \times 5 \textrm{ GHz}$, $\kappa_i = 2\pi \times 100 \textrm{ MHz}$, $g_1(0) = 0.2\kappa_\mathrm{ex} = g_2(0)$, $\alpha_1(0) = 0.7435$, $\alpha_2(0) = 0.5470$, and $\beta_{\alpha}(0) = 0.3650$.}
	\label{fig:10i0.2}
\end{figure*}
One issue to be addressed is the physically valid range of $g_1(t)$ and $g_2(t)$. For the example of optomechanical oscillator blocks, if we store the phonon memory in a membrane at an antinode of the cavity wave and then displace the equilibrium position using two laser drives highly detuned from the cavity field, then the coupling coefficient will carry different signs for opposite displacements \cite{karuza2012tunable, thompson2008strong}. For generic memory blocks, if we set $|g_1(0)| < |\kappa_{\mathrm{ex},1} + \kappa_i|/4$ and $|g_2(0)| < |\epsilon\kappa_{\mathrm{ex},1} - \kappa_i|/4$, then the constants $C$ and $D'$ in the coefficient expressions become real, thereby replacing the oscillatory behavior of the coupling profiles with a superposition of exponentials. In that case, we can maintain fully positive values for $g_1(t)$ and $g_2(t)$ for select values of $\alpha_1(0)$, $\alpha_2(0)$, and $\beta_{\alpha}(0)$. Figure~\ref{fig:10i0.2} depicts the time-evolution of the state populations and pulse shapes for $g_1(0) = 0.2\kappa_{\mathrm{ex},1} = g_2(0)$, $\alpha_1(0) = 0.7435$, $\alpha_2(0) = 0.5470$, and $\beta_{\alpha}(0) = 0.3650$. Here, we see that the Rabi oscillations have disappeared, and $g_1(t)$ and $g_2(t)$ asymptotically decay to 0.
\par
We also made an argument earlier that since the system starts with 0 or 1 excitation, the total number of excitations in the system is never greater than 1, which enabled us to restrict the Hilbert space to just the vacuum state and the singly-excited states.
\begin{figure*}[!tb]
	\centering
	\subfloat{\includegraphics[width=0.48\textwidth]{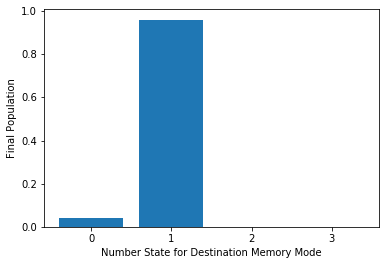}}
	\subfloat{\includegraphics[width=0.48\textwidth]{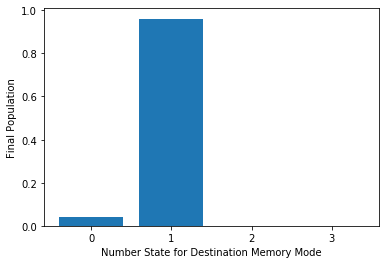}}
	\caption{Final occupation probabilities for $b_2$ number states for $\kappa_\mathrm{ex} = \kappa_{\mathrm{ex},1} = \kappa_{\mathrm{ex},2} = 2\pi \times 5 \textrm{ GHz}$, $\kappa_i = 2\pi \times 100 \textrm{ MHz}$, $g_1(0) = \kappa_\mathrm{ex} = g_2(0)$, $\alpha_1(0) = 0.7$, $\alpha_2(0) = -0.7$, and $\beta_{\alpha}(0) = 0$ (a), and $\kappa_\mathrm{ex} = \kappa_{\mathrm{ex},1} = \kappa_{\mathrm{ex},2} = 2\pi \times 5 \textrm{ GHz}$, $\kappa_i = 2\pi \times 100 \textrm{ MHz}$, $g_1(0) = 0.2\kappa_\mathrm{ex} = g_2(0)$, $\alpha_1(0) = 0.7435$, $\alpha_2(0) = 0.5470$, and $\beta_{\alpha}(0) = 0.3650$ (b).}
	\label{fig:population tomograph}
\end{figure*}
Figure~\ref{fig:population tomograph}, based on a full-quantum simulation of the density matrix evolution performed in QuTiP (Quantum Tools in Python), numerically corroborates this analysis, as the final occupation probability for the states $\ket{2}$ and $\ket{3}$ in the $b_2$ mode is zero. Furthermore, the simulations confirm the 96\% maximal fidelity value calculated for the optomechanical oscillator, since a transfer process that starts at $\ket{1}$ in the $b_1$ mode ends with an occupation probability of 0.96 for $\ket{1}$ in the $b_2$ mode, with a probability of 0.04 that the system collapses to vacuum due to the loss channels.
\par
Lastly, we make a brief note about the meaning of the time arguments for the coupling profiles and state coefficient functions. Since the two blocks are spatially separated, the intermediate state takes some time $\tau$ to travel from S1 to S2, and therefore the clock for the S2 functions will lag that for the S1 functions by $\tau$.

\section{Conclusions}

We derived the analytical expressions for the coupling coefficient control sequences that accomplish the optimal quantum state transfer fidelity for the most general model and parameters of two disparate memory blocks. The optimality is verified numerically through comprehensive simulations, resulting in 96\% quantum fidelity of state transfer with practical system parameters for optomechanical memory blocks. We also studied the effect of the nonidealities - the intrinsic losses and the asymmetric parameters between the two blocks. The fidelity exhibits degradation over the intrinsic losses, resembling an exponential decay tendency. The obtained analytical expression for the coupling controls is widely applicable in any practical integrated quantum memory blocks with extremely long decoherence times and tunable coupling rates. We are currently in the process of building an integrated system for further experimental verification.

It is noteworthy that our theory is applicable to an addressable quantum memory stack with multiple memory blocks. For example, the quantum information of S1 memory block can be transferred to a desired address of quantum memory through the coherent switching of the output $a_{\mathrm{out},1}$ to an input $a_{\mathrm{in}, n}$ of Sn memory block via the programmable nanophotonic processors \cite{harris2017quantum, zhuang2015programmable}. Memory blocks may have different energy gaps between the two qubit basis states if the coupling between $a_n$ and $b_n$ fields is achieved through a parametric process where the mediating classical signal embedded in $g_n(t)$ may have the oscillation frequency to cancel the frequency difference between $a_n$ and $b_n$ fields. 

This work is supported by the U.S. Department of Energy, Office of Science, Advanced Scientific Computing Research (ASCR) under FWP 19-022266 ‘Quantum Transduction and Buffering Between Microwave Quantum Information Systems and Flying Optical Photons in Fibers’. The work was also supported by the Laboratory Directed Research and Development program at Sandia National Laboratories, a multimission laboratory managed and operated by National Technology and Engineering Solutions of Sandia, LLC., a wholly owned subsidiary of Honeywell International, Inc., for the U.S. Department of Energy’s National Nuclear Security Administration under Contract No. DE-NA-003525. This paper describes objective technical results and analysis. Any subjective views or opinions that might be expressed in the paper do not necessarily represent the views of the U.S. Department of Energy or the United States Government. M.E. performed this work, in part, at the Center for Integrated Nanotechnologies, an Office of Science User Facility operated for the U.S. Department of Energy (DOE) Office of Science.

\bibliography{bib-ref}
\bibliographystyle{apsrev4-1}

\end{document}